\documentclass[conference]{IEEEtran}

\usepackage{graphicx}
\usepackage{xcolor}
\usepackage{amsfonts}
\usepackage{mathtools}
\usepackage{algpseudocode}
\usepackage{float}
\usepackage{nomencl}
\usepackage{cite}
\usepackage{amsmath}
\usepackage{amssymb}
\usepackage{amsfonts}
\usepackage{textcomp}
\usepackage[linesnumbered,ruled,vlined]{algorithm2e}
\usepackage{xspace}

\def\BibTeX{{\rm B\kern-.05em{\sc i\kern-.025em b}\kern-.08em
    T\kern-.1667em\lower.7ex\hbox{E}\kern-.125emX}}

\begin{document}

\title{Understanding how social discussion platforms like Reddit are influencing financial behavior}

\author{\IEEEauthorblockN{Sachin Thukral}
\IEEEauthorblockA{\textit{TCS Research} \\
Delhi, India \\
sachi.2@tcs.com}
\and
\IEEEauthorblockN{Suyash Sangwan}
\IEEEauthorblockA{\textit{TCS Research} \\
Delhi, India  \\
suyash.sangwan@tcs.com
} \\ 
\IEEEauthorblockN{Pramit Kumar Chandra}
\IEEEauthorblockA{\textit{IIT Kharagpur} \\
West Bengal, India \\
dchandra13165@gmail.com
}
\and
\IEEEauthorblockN{Arnab Chatterjee}
\IEEEauthorblockA{\textit{TCS Research} \\
Delhi, India \\
arnab.chatterjee4@tcs.com
}
\and
\IEEEauthorblockN{Lipika Dey}
\IEEEauthorblockA{\textit{TCS Research} \\
Delhi, India \\
lipika.dey@tcs.com
} \\
\IEEEauthorblockN{Animesh Mukherjee}
\IEEEauthorblockA{\textit{IIT Kharagpur} \\
West Bengal, India \\
animeshm@gmail.com
}
\and
\IEEEauthorblockN{Aaditya Agrawal}
\IEEEauthorblockA{\textit{IIT Kharagpur} \\
West Bengal, India \\
aaditya1106@gmail.com
}
}

\maketitle

\newcommand{\ps}[1]{\textcolor{green}{[Punyajoy: #1]}}
\newcommand{\st}[1]{\textcolor{blue}{[Sachin: #1]}}
\newcommand{\ac}[1]{\textcolor{magenta}{[Arnab: #1]}}
\newcommand{\ag}[1]{\textcolor{cyan}{[Aaditya: #1]}}
\newcommand{\am}[1]{\textcolor{red}{[AM: #1]}}

\begin{abstract}
This study proposes content and interaction analysis techniques for a large repository created from social media content. Though we have presented our study for a large platform dedicated to discussions around financial topics, the proposed methods are generic and applicable to all platforms. Along with an extension of topic extraction method using Latent Dirichlet Allocation, we propose a few measures to assess user participation, influence and topic affinities specifically. Our study also maps user-generated content to components of behavioral finance. While these types of information are usually gathered through surveys, it is obvious that large scale data analysis from social media can reveal many potentially unknown or rare insights. Characterising users based on their platform behavior to provide critical insights about how communities are formed and trust is established in these platforms using graphical analysis is also studied.
\end{abstract}

\begin{IEEEkeywords}
social network analysis, behavior, finance, Reddit
\end{IEEEkeywords}

\section{Introduction}
\label{intro}
As the financial industry embraces major changes brought in by rapid digitization, social media is emerging as a necessary channel to reach younger clients. The millennials who are mostly employed now or have some income, are looking for savings options as well as seeking other financial advice from dedicated social media platforms. Based on surveys conducted on students and young professionals, several studies also confirm that social media is playing an increasingly important role in financial decision-making \cite{Yanto, Isomidinova}. These surveys were designed to cover five variables, namely financial behavior, financial attitude, financial knowledge, social media exposure, and peer influence of the subjects.
The responses confirm that social media plays a strategic role in shaping the financial behavior of the millennials since that is where they are looking for general financial knowledge, and are also seeking advice for planning and executing financial activities. Interestingly, as per one site, as much as 70\% of Gen Z are also searching for retirement plans\footnote{https://www.putnam.com/advisor/content/advisorTechTips/3536-for-gen-z-tech-is-money}. Thus, it is quite clear that these platforms are playing a big role in imparting financial literacy and shaping financial attitudes of society. Social media also acts as an aggregator of information, reducing the cost and effort of sourcing knowledge from all over the internet.

Since surveys are restrictive, in this paper, we explore the use of natural language processing techniques and social network analysis to analyze large volumes of discussions and interactions that are getting created in financial discussion platforms. We have chosen Reddit for our study because it has emerged as a premier platform where netizens engage in serious and active discussions on contemporary topics. Reddit discussions are carefully moderated. While the active members voice their opinions and share experiences and information, many active but passive subscribers consume information from Reddit. Hence, Reddit is considered one of the key information dissemination platforms. However, the proposed methods can be used for any other platform also. Users participate in such platforms in one or more of the following ways:
\begin{itemize}
    \item Initiate a discussion by posting some content
    \item Respond to other users by commenting on their posts
    \item Engage with other users by continuing on the comment thread.   
\end{itemize}
 In this paper, we have proposed methods that are designed to extract insights about the topics of conversation and the interaction patterns among the users on these platforms. Along with analyzing the content, which reveals the topics of discussion, this paper also proposes various measures that can be used to characterize the social behavior of users on this platform. We propose measures to compute user participation and influence in the platform, and thereby automatically identify the highly influential users and their circles of influence. We have shown that the proposed methods and measures can not only reveal information about the financial needs and aspirations of people, but also help in ascertaining the maturity of financial knowledge of the platform users, identifying communities, and also reveal some interesting insights about the relationships between individuals and communities. 
 
The key contributions of the paper are summarized as follows:
\begin{itemize}
\item [1.] Proposes content and interaction analysis of a large repository created from a social platform dedicated to discussion around financial matters. A novel method has been proposed to extend topic extraction using Latent Dirichlet Allocation (LDA) to detect the optimal number of topics from a large repository, such that the maximum number of posts can be assigned to a key topic.   
\item [2.] Maps user-generated content to components of behavioral finance. The acquired knowledge offers interesting insights about users in terms of their key financial concerns, risk-taking potential, entrepreneurial attitude, etc. While these types of information are usually gathered through surveys, it is obvious that large-scale data analysis from social media can reveal many potentially unknown or rare insights.
\item [3.]	Characterizes users based on their platform participation and interactions to identify the high-influence users automatically.
\item [4.]	Provides insights about how communities are formed and trust is established in these platforms using graphical analysis.
\end{itemize}

The rest of the paper is organized as follows. Section~\ref{sec:literacy} provides a brief introduction to financial literacy as defined in the literature. Section~\ref{sec:dataset} introduces the dataset we have used to derive insights. Section~\ref{sec:behavior} introduces the measures defined for assessing behavior. Section~\ref{sec:topicmodeling} describes how content of posts are analyzed. Section~\ref{sec:network} presents network analysis tasks undertaken.  

\section{Brief introduction to Financial Literacy}
\label{sec:literacy}
Several studies have shown that literate people are better at budgeting, saving, spending, debt management, participating in the stock market, planning pension funds, etc. In other words, the higher an individual’s knowledge and understanding of financial concepts, the more likely they are to behave financially well in their daily life \cite{andarsari2019role}. However, it is a fact that the level of financial literacy varies a lot across a society. Of late, there has been an active initiative to improve the financial literacy of citizens to ensure responsible financial behavior.  Financial literacy empowers a person to make smart financial choices. 
According to the Financial Literacy and Education Commission, US Treasury department\footnote{https://www.federalreserve.gov/2015-report-economic-well-being-us-households-201605.pdf}, the five key components of financial literacy are defined around the following financial activities:
\begin{itemize}
\item	Earn – which is targeted at understanding one’s paycheck, essentially about the earnings and the deductions. 
\item	Spend – this aspect is about learning to create a personal budget which reflects a plan around how to spend money within the limits of a budget while achieving personal goals.
\item	Save – this is the most difficult yet most important aspect as it teaches about managing financial goals for the future including retirement plans, catering to medical emergencies, as well as planning for big purchases like a car or house.
\item	Borrow - Financial literacy teaches that borrowing is not essentially bad provided one is a diligent saver and does it knowledgeably. Maintaining a good credit score, and keeping track of products and services with good interest rates are keys to intelligent borrowing. 
\item	Protect - Learning to protect one’s money from fraud, scams, and malpractices is the fifth pillar of financial security. 
\end{itemize}

Financial literacy shapes financial behavior and thereby influences the financial activities that a person undertakes. Financial literacy has been found to influence the financial decisions of people in many direct and indirect ways. Knowledge imparts confidence and thereby helps a person in taking the right decisions while seeking loans, choosing plans, or seeking funding for business opportunities. Studies show that financially literate people are more likely to get access to external funding and to develop their businesses, than those lacking knowledge.

The knowledge source plays an important role in this. While earlier there was high dependence on acquiring knowledge from individuals, social media is fast replacing it as a source where one can obtain the wisdom of many. Reliance on peers is also found to increase. Hence, both individual and collective financial behaviors are also seeing a shift.

\section{Dataset}
\label{sec:dataset}

The social news aggregation, web content rating, and discussion website \textit{Reddit} is organized into well-defined communities, known as \textit{subreddits}, where members take part in discussions around a common topic by creating posts or creating comments to posts or earlier comments. For this work, users who have made at least one post or comment are considered for analysis, interchangeably referred to as users or authors.

We used the \textit{Pushshift API}~\cite{Reddit_dump}, to crawl the data from Reddit, and we have gathered $1,34,521$ posts and $15,21,940$ comments from finance related subreddit \textit{r/personalfinance} from 1st July 2020 to 30th June 2021 which is hereafter referred to as the \textbf{Finance Dataset}. We chose this platform due to its large subscriber base which is around 15 million. It shows that the number of comments on this platform is far more than the number of posts, which indicates that users actively engage with other peer subscribers over content created by the latter. 
This is further confirmed by the fact that while 18\% of users have made more than 5 comments over the period, only  1\% of users have made more than 5 posts.
This is a clear indicator that users find the responses engaging, and may not be posting original content since the existing posts and comments already satisfy their requirements.    

In the next section, we provide more details about the platform's activities.

\section{Platform activity statistics}
\label{sec:behavior}
As mentioned in the earlier section, the number of comments was found to far outnumber the number of original posts.  Delving deeper into the actions revealed that the average number of comments per post is 11, and only 5\% of the posts had received no comments. This indicates that for an overwhelming majority of the posts, users engage actively with each other over comments, and very few posts remain unnoticed.

\begin{table}[h]
\caption{Author Categories based on their activity}
	\label{table:author_categories}
	\begin{center}
	    \begin{tabular}{|p{1.7cm}|r|r|r|}
	    \hline
	    Author Category & Authors(\%) & Posts(\%) & Comments(\%) \\
	    \hline
	    Only Posting & 12.88 & 25.62 & 0  \\
        \hline
        Only Commenting & 59.11 & 0 & 51.9 \\
        \hline
        Both & 28.01 & 74.38 & 48.1 \\
        \hline
        \end{tabular}
	\end{center}
\end{table}

Table~\ref{table:author_categories} presents activity statistics of the users of this platform. We see that there are three distinct categories of users. 59\% of the users are only commenting. 13\% of the subscribers are only posting content, but not commenting on other's posts. Remaining 28\% of the users are posting as well as commenting. It was also observed that out of those who have authored new posts, $20\%$ users have created unique posts more than once.
However, when it comes to commenting, it is observed that as high as $57\%$ of the users have commented more than once, which re-emphasizes that lot of users are actively engaged with the community, and engagement is not necessarily happening through creation of unique posts.
The next subsection reveals some more insights about response time and quality of comments.
 
Users on a subreddit platform cast votes to share their appreciation for content. Every post or comment is awarded one vote by the system. The final score of a post or a comment is the aggregate of votes received from other users. In the current dataset, around $92\%$ of the posts and $72\%$ of comments remained at score 1. 

\begin{figure}[h]
    \centering
    \includegraphics[width=\linewidth]{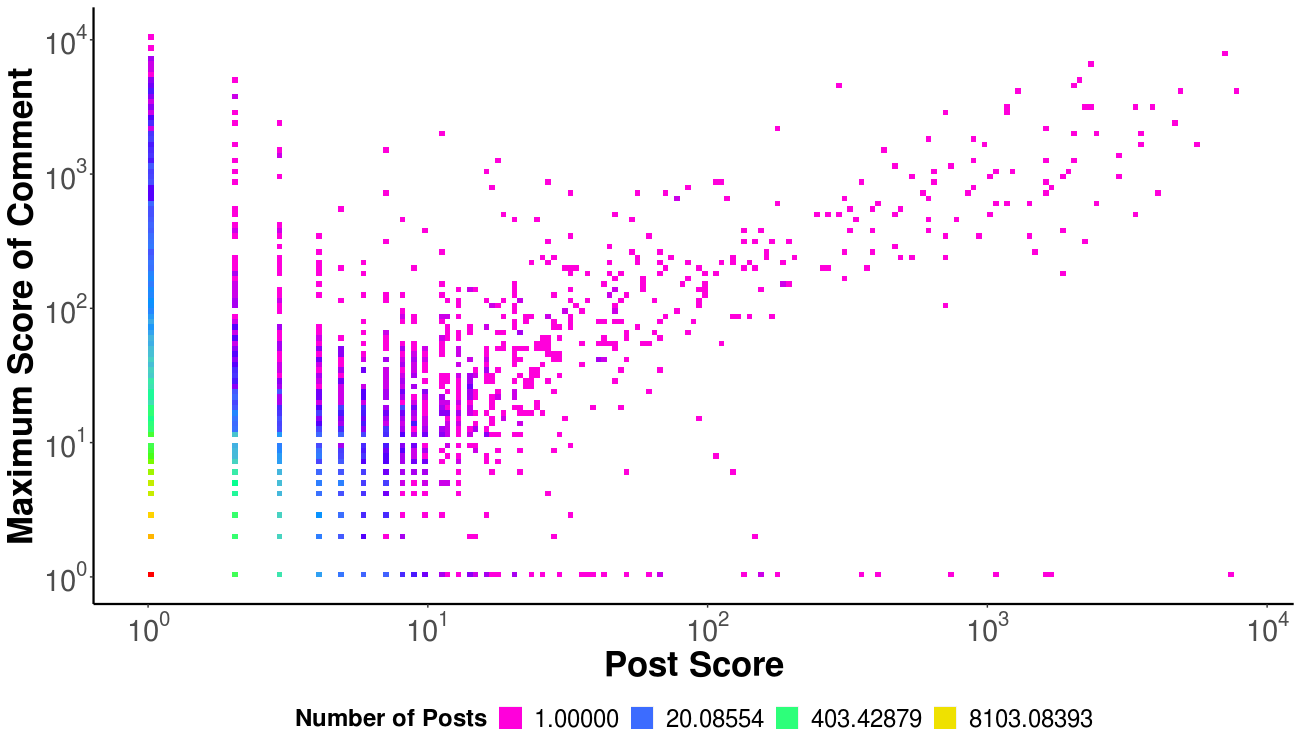}
    \caption{Posts score vs. maximum score a comment received on the post [Both axes are log10 scaled]}
    \label{fig:score_max}
\end{figure}

In order to study the relationship between post score and comment scores, for each post we obtained the maximum score received by any comment in its comment thread. Fig.~\ref{fig:score_max} shows the plot of post scores vs their maximum comment scores.  There are two significant observations:
\begin{itemize}
    \item Many low scoring posts have comments which have scored very high. These are manifested as vertical lines present for the low scoring posts closer to the origin along the x-axis. This indicates that even if the users have not voted for the original post, the comments made on it have been highly appreciated. In other words, attraction is garnered not only by posts, but by the comments received also. It may be inferred that, many users might have had the same query or concern as expressed in the original post, and therefore found the answer to be useful. Thus the score received by the comments can act as a proxy measure for judging the utility of the platform.  
    \item High scoring posts tend to have comments that are also high scoring, which is means the content as a whole is appreciated. 
\end{itemize} 
The above observations are also consistent with the earlier observation that comments play a very significant role in this platform. 

\begin{figure}[h]
    \centering
    \includegraphics[width=\linewidth]{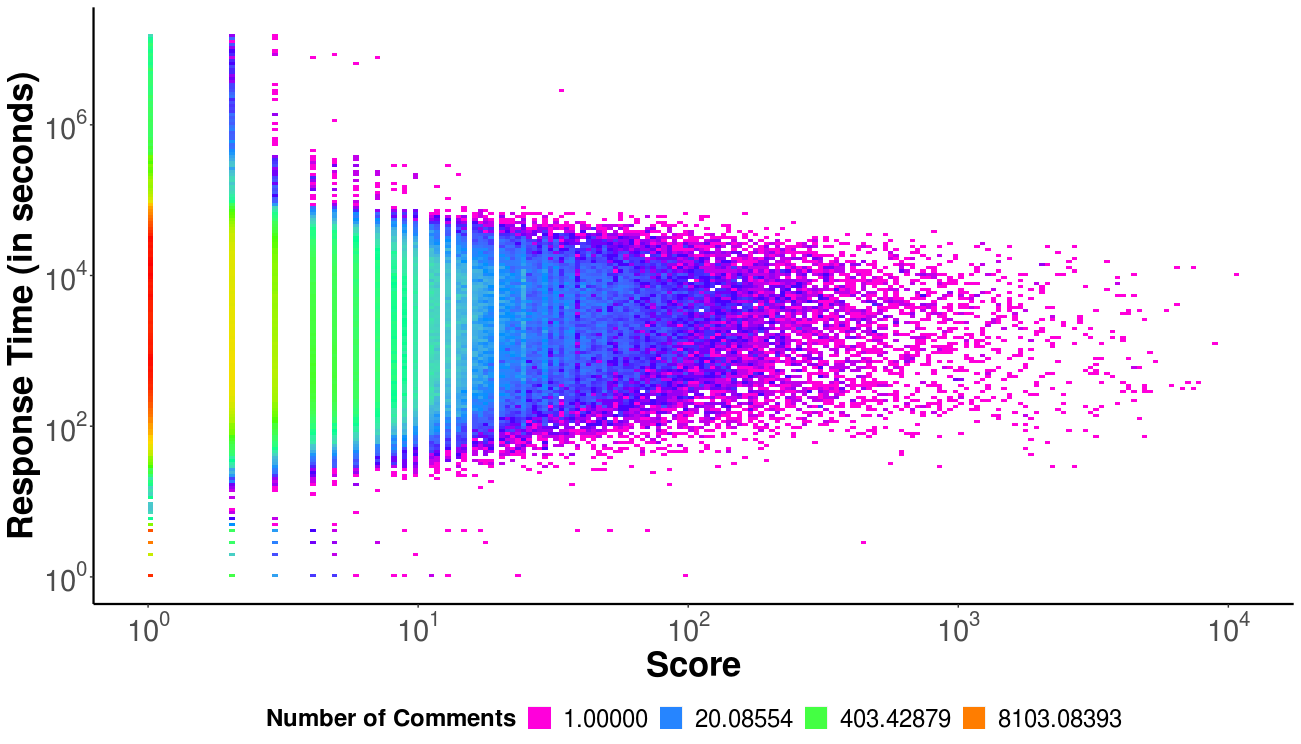}
    \caption{Comment Score vs. Response time(s) [Both axes are log10 scaled]}
    \label{fig:response_time}
\end{figure}

We have defined \textit{response time} as the time lapse between a comment getting posted in response to its parent, which can be either a post or another comment. Fig.~\ref{fig:response_time} shows for all the comments, the density plot of response time versus the score received by a comment. The graph is plotted using log scales on both the axes. Each point on the plot denotes a collection of comments bunched together, all of which have the same response times and scores. The colour of a point is decided by the cardinality of the bunch, where the colour scale moves from violet to red, as the cardinality goes from low to high. The concentration of blue points show that the response time of majority of the comments varies from 32 seconds to 9 hours. It is observed that on an average, a high scoring (score $> 100$) response is received within two hours. The left hand side of the graph shows that many posts have received comments even after two months. Clearly, it can be inferred that users find the platform responsive and therefore useful.   

\section{Content analysis of Financial Subreddit using Topic Modeling}
\label{sec:topicmodeling}

Unsupervised models like topic extraction help in obtaining a bird’s eye view of the content of a large repository. We have used the Latent Dirichlet Allocation Machine Learning for Language Toolkit (LDA MALLET) for extracting topics\cite{rehurek2011gensim} from the posts. Given a fixed number of topics, say $k$, the topics are obtained using an optimized Gibbs sampling algorithm\cite{yao2009efficient} which calculates the optimized probability of each word in a document to belong to a particular topic, and thereafter the distribution of all $k$ topics in each post. Each topic is represented by the top $n$ words that have the highest probability of belonging to it. One of the key challenges of the above method is to determine the optimal number of topics for a repository. LDA MALLET offers two intrinsic measures to determine the topics - \textit{perplexity measure} and \textit{coherence score}. \textit{Perplexity} captures how surprised a model is by new data it has not seen before and is measured as the normalized log-likelihood of a held-out test set. This did not suit our purpose. \textit{Coherence score}, on the other hand, is computed by measuring the degree of semantic similarity between high scoring words of a topic, and selects a number of topic that makes each one most coherent. We chose this measure initially, but found that a large number of posts ended up receiving a uniform distribution or near equal probabilities for all topics, which made topic interpretation difficult and noisy. This led us to come up with a new measure that is defined in terms of \textit{skewness} of topic distribution for each document to determine the optimal number of topics.

For each post $P_i$, let $T_1$, $T_2$, .... $T_k$ be the topical distribution of all $k$ topics given by LDA MALLET, and $\mu(P_i)$, $\sigma(P_i)$ and $\cal{M}(P_i)$ be the mean, standard deviation and median respectively, derived from it.

`Skewness' is a measure of the asymmetry of the probability distribution of a real-valued random variable around its mean, and is defined as:

\begin{align*}
     {\rm Skewness}(P_i) = \frac{3(\mu(P_i) - {\cal M}(P_i))}{\sigma(P_i)}.
\end{align*}

The skewness value can be positive, zero, negative, or undefined. Positive skew value indicates that average topical distribution is greater than the median value of the distribution, which further implies that at least one of the topics is significantly present. A zero or negative skew on the other hand indicates that none of the topics is significantly present in the post. If $k$ denotes the number of topics, let $W_k$ denote the count of posts with negative skewness measure.
In our implementation, starting with an arbitrary value of $k$, the intent is to minimize $W_k$, by increasing $k$ iteratively, till no further significant improvement in $W_k$ is noticed. We set initial $k$ to 2, though it could be higher. Algorithm~\ref{alg:1} presents the algorithm for this.      

\begin{algorithm}[h]
    \caption{Extracting Optimal Topics}
    \label{alg:1}
    \SetAlgoLined
    \SetKwInOut{Input}{input}
    \SetKwInOut{Output}{output}
    \DontPrintSemicolon
    \Input{Set P of posts}
    \Output{k topics \Comment{Optimal number of Topics}}
    
       //Initialization
       
    \quad \quad \quad $MinSkew \gets inf$
    
    \quad \quad \quad $k \gets 2$
    
    \For  {$i \in 2 \to N$} {
        $Model \gets LDA\_MALLET(P, i)$,
        
        $NewSkew  \gets 0;$ 
        
        \For {$each\ post\ p \in P$} {  
        
        Compute skewness(p);
        
            \If {$skewness(p) <= 0$}{
            
                $NewSkew++$
            }
        }
        \If {$NewSkew < MinSkew$}{
            
            $MinSkew \gets NewSkew$
    
            $k \gets i$
        }
    }
\Return $(k, Model(k))$
\end{algorithm}

Implementing it over the current set of 134521 posts, the optimal number of topics was found to be 11, where the cardinality of $W_{11}$ settled at 404, which is only 0.3\% of the original posts. The top 10 words of each topic ($T_j$) (where $j=1,\ldots,k$) are shown in Table~\ref{table:seed_words_finance}. The names are manually assigned based on the top words. As can be seen from the topics words, they can be easily mapped to the key financial attributes.
Top 20 posts with highest topic probabilities for each topic were also manually examined to verify the mapping :
\begin{itemize}
    \item \textbf{Earn}- This attribute is most reflected for posts that have the fourth topic with words like job, offers, salary, positions and hike.
    \item \textbf{Spend}- Topics like `Credit Cards', `Vehicle Purchase and Insurance', and 'Housing related' can be mapped to this component where people are seeking advice while buying car, insurance and house.
    \item \textbf{Save}- Posts coming under `Stocks Investment', `Annual Tax', and `Savings' can be linked to this attribute. Here posts are more naive where authors have queries like `how to save', `how much should i spend on travel', etc. Moreover, top topic words of `saving' topic are like `college', `school', `students' which depict that their demographic attribute is mainly young people.
    \item \textbf{Borrow}- `Loan' topic is related to borrowing component as people are looking for refinance, less interest rates.
    \item \textbf{Protect}- The posts that come under the topic `Complaints about banking operations' can be linked to  malpractices. 
\end{itemize}
Two topics namely `Family related Finances' and `Advices and Questions' could not be directly mapped to any specific financial attribute. However, as we discuss later, these topics often co-occur with other topics, and thus provide a lot of context for the core issues expressed in the post.
\begin{table}[h]
\caption{Topicwise top words}
	\label{table:seed_words_finance}
	\begin{center}
	    \begin{tabular}{|p{2cm}|p{3cm}|p{2.5cm}|}
	    \hline
	    \textbf{Topic Name} & \textbf{Top 10 Words}  & \textbf{Financial attributes} \\
	    \hline
	    Credit Cards & 'pay', 'credit', 'debt', 'card', 'bill', 'payment', 'credit score', 'payment', 'limit', 'collection' & Spend  \\
        \hline
        Family Related Finances & 'money', 'parent', 'live', 'family', 'life', 'lose', 'time', 'mom', 'friend', 'leave'  & - \\
        \hline
        Vehicle Purchase and Insurance & 'car', 'insurance', 'buy', 'month', 'finance', 'purchase', 'lease', 'cost', 'vehicle', 'sell' & Spend \\
        \hline
        Job & 'work', 'job', 'company', 'offer', 'start', 'week', 'salary', 'year', 'business', 'pay' & Earn \\
        \hline
        Stocks Investment & 'account', 'money', 'invest', 'fund', 'investment', 'stock', 'open', 'cash', 'retirement', 'transfer'  & Save \\
        \hline
        Loan & 'loan', 'pay', 'year', 'mortgage', 'payment', 'interest', 'rate', 'low', 'refinance', 'month' & Borrow \\
        \hline
        Advice and Questions & 'good', 'question', 'advice', 'time', 'thing', 'post', 'people', 'read', 'advance', 'understand' & - \\
        \hline
        Complaints about Banking Operations & 'bank', 'account', 'check', 'call', 'receive', 'send', 'charge', 'back', 'number', 'refund' & Protect \\
        \hline
        Housing Related & 'house', 'home', 'move', 'buy', 'rent', 'live', 'year', 'sell', 'place', apartment' & Spend \\
        \hline
        Annual Tax & 'year', tax', 'income', 'plan', 'contribute', 'state', 'file', 'employer', 'retirement', 'question' & Save \\
        \hline
        Savings & 'saving', 'year', 'money', 'start', 'debt', 'plan', 'expense', 'college', 'school', 'advice' & Save \\
        \hline
		\end{tabular}
	\end{center}
\end{table}

In order to detect the dominant topics of a post, for each topic $T_i$, we computed mean ($\mu_i$) and standard deviation($\sigma_i$) of the probability values of the topic across all posts. For each post, if topic probability of $T_i$ is greater than $\mu_i + \sigma_i$, then topic $T_i$ is assumed to be a dominant topic of the post. Each post may have more than one dominant topic. This further ensures that while we do a quantitative analysis of the topic presence in the repository, only the posts for which the topic is dominant are taken into account, and its chance or noisy presence is ignored. Based on number of posts for which a topic is dominant, we compute the distribution of topics over the entire repository. This is shown in Table~\ref{table:finance_topic_statistics}. The presence of the topics are fairly uniform. Only 0.94\% of the posts could not be assigned any dominant topic. Around 50\% of the posts had one dominant topic, while the rest had more than one topic. 

\begin{table}[h]
\caption{Topicwise Distribution}
	\label{table:finance_topic_statistics}
	\begin{center}
	    \begin{tabular}{|l|r|}
	    \hline
	   \textbf{ Topic Name } & \textbf{Presence of Topic (\%)}  \\
	    \hline
	    Credit Cards & $13.88$   \\
        \hline
        Family Related Finances & $12.36$ \\
        \hline
        Vehicle Purchase and Insurance & $10.89$ \\
        \hline
        Job  & $11.85$ \\
        \hline
        Stocks Investment & $19.36$ \\
        \hline
        Loan & $13.46$ \\
        \hline
        Advice and Questions & $11.13$ \\
        \hline
        Complaints about Banking Operations & $16.87$  \\
        \hline
        Housing Related & $13.61$ \\
        \hline
        Annual Tax & $17.50$ \\
        \hline
        Savings & $15.68$ \\
        \hline
		\end{tabular}
	\end{center}
\end{table}

\begin{figure}[h]
    \centering
    \includegraphics[width=\linewidth]{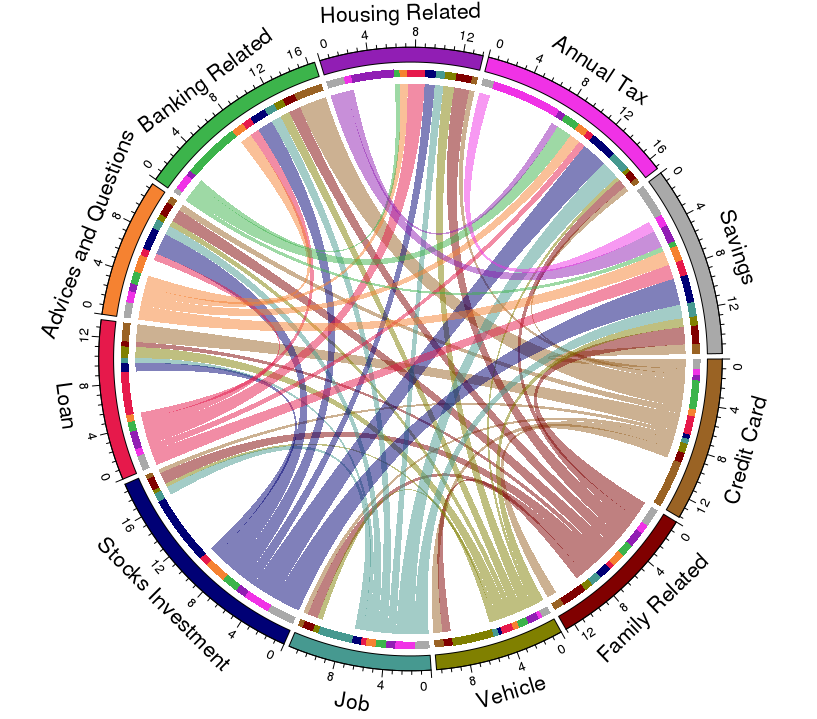}
    \caption{Topic Co-occurences}
    \label{fig:topic_cooc}
\end{figure}

Fig.~\ref{fig:topic_cooc} provides insights about co-occurring topics. Each arc represents a topic, while chords represent topic co-occurrence of a pair of topics. The white portion of each arc represents occurrence of that topic only. Fig.~\ref{fig:topic_cooc} shows that "Credit Card" and "Complaints about Banking" related topics co-occur most often. "Stocks Investment" and "Savings" also co-occur frequently. 
It may be noted that the topics "Advice and Questions" and "Family related finance" have the least white space area attached to them, and co-occur frequently with all other topics. This confirms the fact that the users of this platform are actively seeking help for all financial tasks. Judging from the high scores received by the comments, it can also be safely concluded that the users do get useful advice and help, which are liked by all. Thus this platform is indeed advocating financial literacy, that is trusted by many people.  

\begin{figure}[h]
    \centering
    \includegraphics[width=\linewidth]{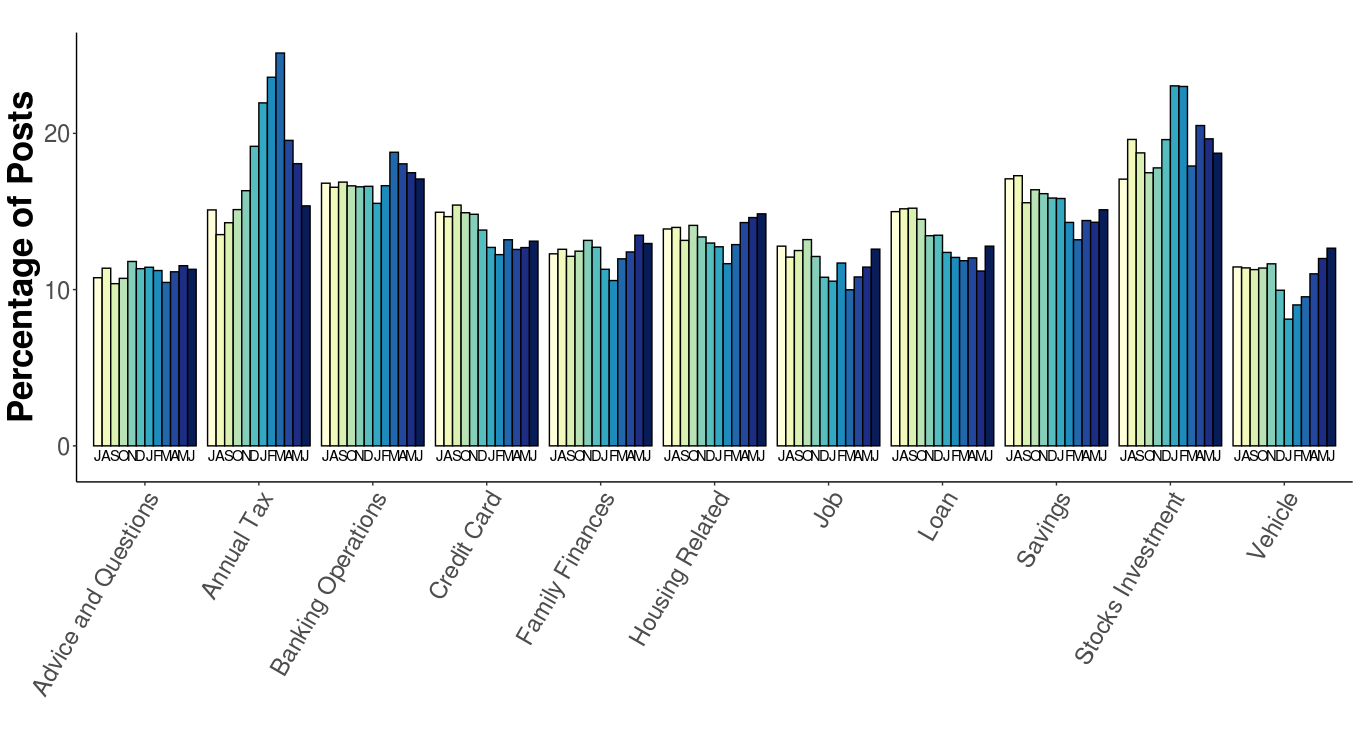}
    \caption{\textit{Monthwise Topic Distribution:} Here for each topic we have bars from the July 2020 to June 2021.}
    \label{fig:monthwise_topic_distribution}
\end{figure}

Fig~\ref{fig:monthwise_topic_distribution} presents month-wise break up of the topics based on their dominant presence. As expected number of posts with topic `Annual Tax' spikes from January to March. Discussion around `Loan' and `Savings' topic is comparatively higher during July to September. All other topics are more or less consistent across the year. 

\subsection{Empath Categorization of Topics}
\begin{figure}[h]
    \centering
    \includegraphics[width=\linewidth]{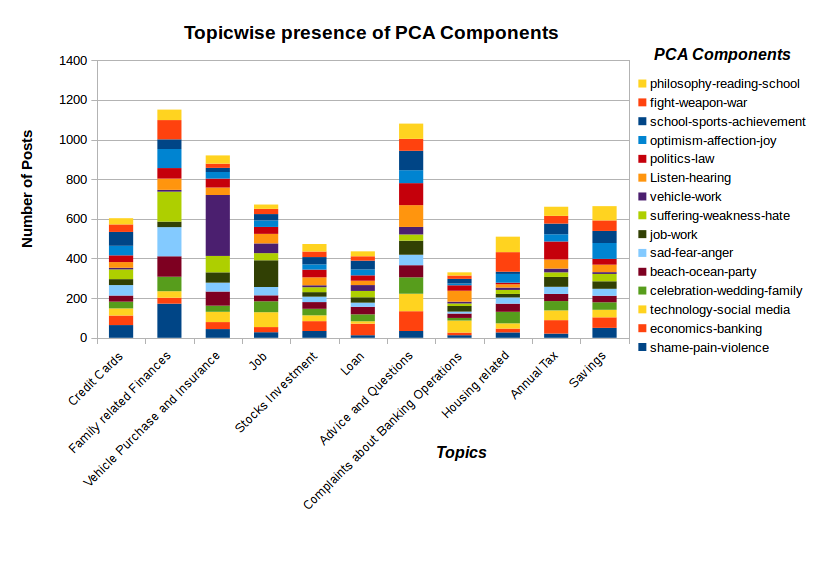}
    \caption{Topicwise Presence of PCA Components}
    \label{fig:Topicwise_presence_of_PCA_Components}
\end{figure}

In order to obtain a deeper perception of the tone of discussions, we used the Empath~\cite{fast2016empath} library which determines the lexical categories present in the content. Empath offers a powerful tool to identify human-interpretable feature sets from text and derive multi-perspective insights. The default library has been trained over a large corpus of 1.8 billion words, to draw connotations between words and phrases, and is capable of detecting 200 pre-trained categories that covers a range of topics from sentiments to professional domains. Each post was analyzed using the Empath tool to determine the categories present in it. The 200 dimensional Empath vectors obtained for the posts were then reduced to fewer dimensions using Principal Component Analysis (PCA). Taking the top 15 components, the key empath categories present in the posts were identified to be as follows : `shame-pain-violence', `economics-banking', `technology-social\_media', `celebration-wedding-family', `beach-ocean-party', `sad-fear-anger', `job-work', `suffering-weakness-hate', `vehicle-work', `listen-hearing', `politics-law', `optimism-affection-joy', `school-sports-achievement', `fight-weapon-war', `philosophy-reading-school'. Fig.~\ref{fig:Topicwise_presence_of_PCA_Components} shows the topic-wise distribution of the empath categories. It was observed that the topic "family related finance" had significant presence of three negative empath categories namely \textit{shame-pain-violence, sad-fear-anger, and suffering-weakness-hate.} As mentioned earlier, "family related finance" co-occurred with other topics. It was found that posts with this topic discussed difficult family situations and thereafter sought help in dealing with financial matters, given such a situation. This is indeed a significant aspect of a social media platform, that users can talk about their family issues uninhibitedly and also seek advice on dealing with it. While it has been reported earlier for other domains, our study shows that this is true for financial advice also. The topic ‘Job’ has comparatively higher presence of work related PCA component. ‘Complaints about banking operations’ has higher presence of technology related Empath components. The topic ‘savings’ showed highest presence of positive Empath categories optimism-affection-joy.

\subsection{Topic Spread of users who comment}

Having obtained insights about activities and topics, we now show how users can be characterized based on their activities and interactions. Posts provide a glimpse into the main content around which discussions take place, and topical analysis showed that most of the posts were to seek help for financial activities. As we saw earlier, a large number of users are only responding to posts and it may be safely assumed that this category of users are providing the guidance and answers to queries. 

We now propose a measure called \textit{topic spread} denoted by $\tau$, to compute topic association for users who respond to posts. This is computed in terms of the dominant topics of the post to which user A responds as follows:  
\begin{align}
\label{eq:concentration}
\tau(A) = \sum_{i=1}^n \frac{1}{s_i^2}
\end{align}
where $s_i$ is the fraction of comments made for posts having topic$_i$ as dominant, and $n$ is total  number of topics. It is observed that around $75\%$ of the authors concentrate on only 1 topic, which means that majority of users have a particular focus within the financial activities. However, users varied widely in their participation. While the average number of comments made by authors was $7.34$, around $7373$ users i.e. around $3.56\%$ have made much more comments. Clearly, these people had more participation and in the next section we propose measures to find more about them.     

\section{Network Analysis combined with Topic insights}
\label{sec:network}

In this section, we attempt to extract  more insights about communities. Unlike standard social network graphs, we create a \textit{directed} graph with users as nodes, and a directed, weighted edge from user A to user B, if A has responded to B's post or comment at least once. 
We first define some network properties here: 
\begin{itemize}
    \item \textbf{Out Degree(A)}: This is the total number of edges going out of A that indicates number of unique users with whom A has interacted.
    \item \textbf{In Degree(A)}: This is the total number of edges going in to A that indicates number of unique users who have responded to A.
\end{itemize}
The graph has $23,334$ nodes and $11,81,108$ weighted edges. Users who have only created posts, but never commented are not a part of this graph. $12.88\%$ users who have out degree 0, are termed as \textit{passive users}, who are not commenting on other's content, but may have received responses from others. $28.54\%$ users have in degree 0 i.e. have not received any explicit response. We now propose a few measures to compute the influence of users.

\begin{figure}[h]
    \centering
    \includegraphics[width=\linewidth]{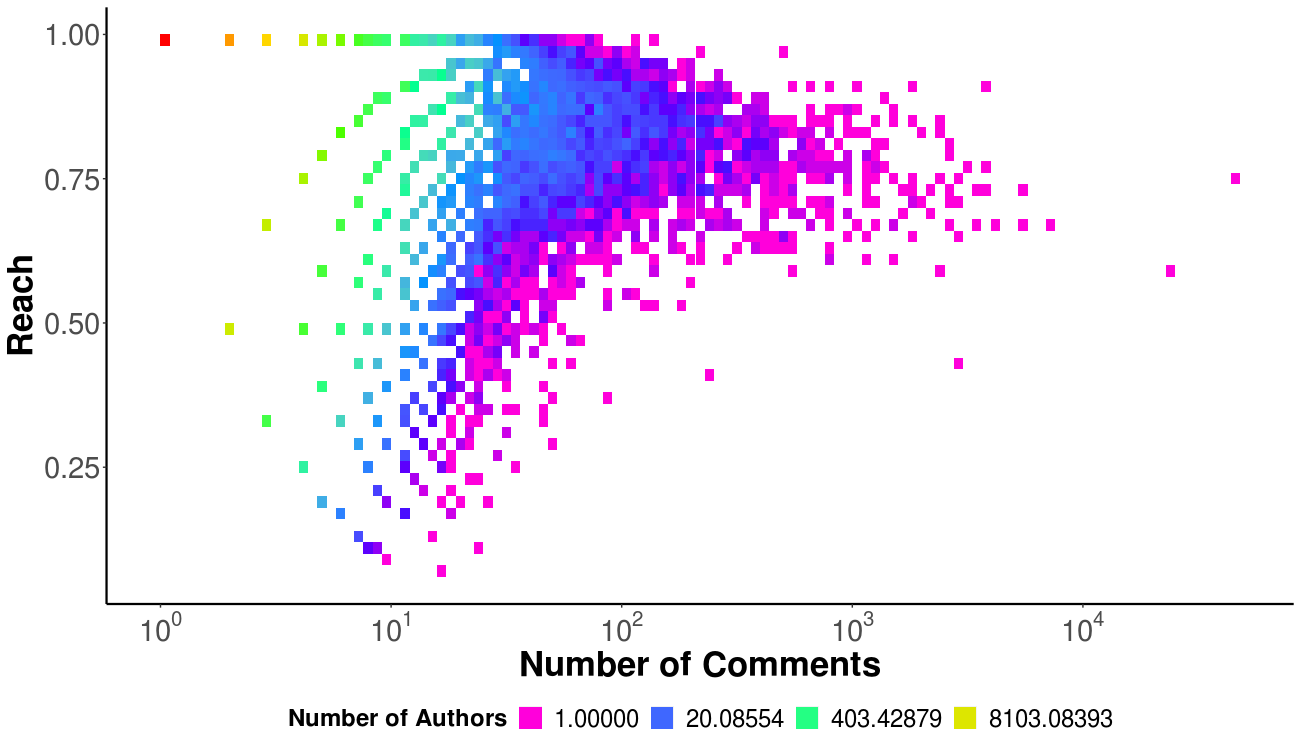}
    \caption{Density Plot for Comments done by author vs their Reach [Here x axes is log10 scaled]}
    \label{fig:reach_author}
\end{figure}

The first measure called $reach$ of a user, is computed as a function of activity and number of unique users one connects to. Denoted by  $R(A)$ for user $A$, it is computed as follows:
\begin{align*}
R(A) = \frac{Out \ Degree(A)}{C(A)}
\end{align*}
where $C(A)$ denotes Total Comments done by A.
Fig~\ref{fig:reach_author} plots the reach of users versus the number of comments done. Most users who have made many comments, have done so for many unique users. We have observed that around $85\%$ commentors have reach greater than 0.75, which means 75\% of their comments are in response to unique authors.  

Next, we compute \textit{contribution} of each user to the forum in terms of percentage contribution to the total number of comments made in the platform. Denoted by $\alpha (A)$ this is computed as  
\begin{align*}
\begin{aligned}
\alpha(A) &= \frac{C(A)}{max(C)}
\end{aligned}
\end{align*}
where $max(C)$ denotes maximum number of Comments Done by any author.
Finally we define a measure called \textit{influence} of a user, denoted by $I(A)$ as
\begin{align}
\label{eq:participation}
I(A) = \alpha(A)*R(A).
\end{align}

\begin{figure}[h]
    \centering
    \includegraphics[width=\linewidth]{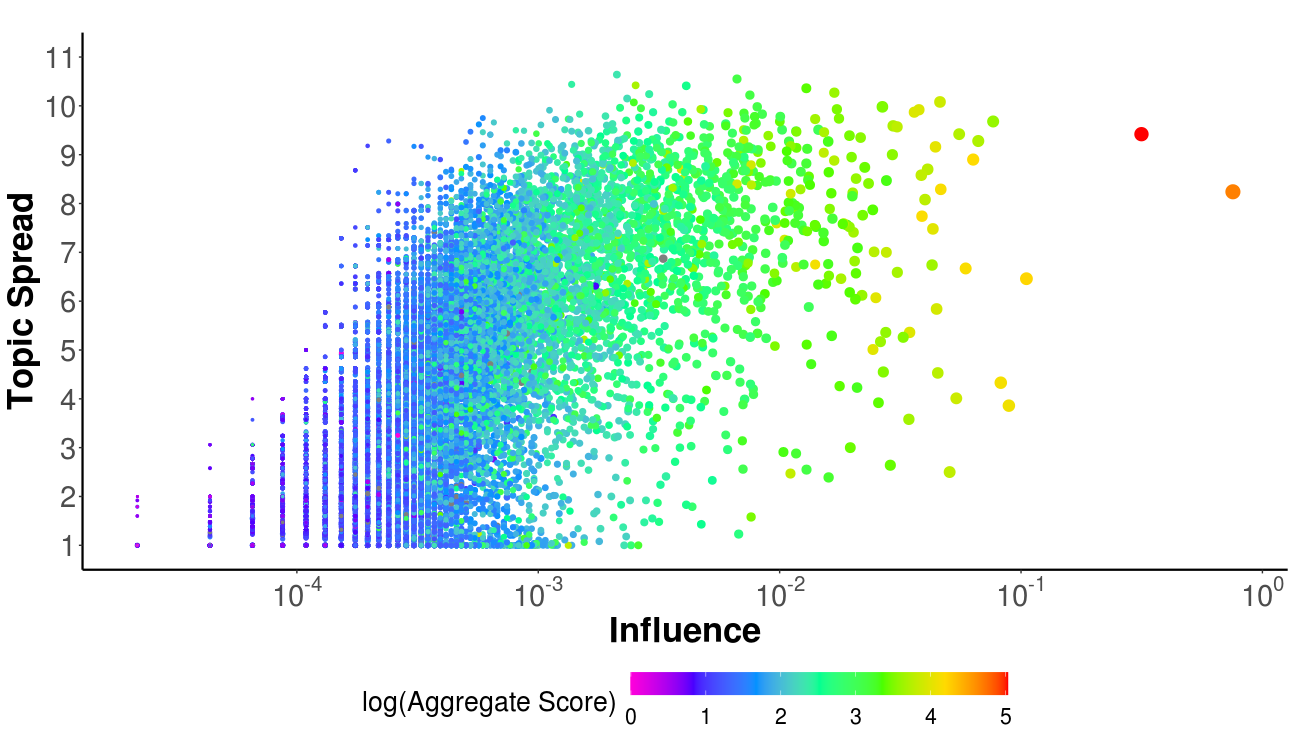}
    \caption{Plot of Topic Spread vs. Influence.
    Each point on the graph represents an author, with the size of the point mapped to the logarithm of the number of comments, and the color mapped to the logarithm of aggregate Comment Score.}
    \label{fig:participation}
\end{figure}

We now present how the three measures \textit{influence, topic spread and scores}, collectively capture the nature of users and interactions on the platform. This is visually captured in Fig.~\ref{fig:participation}. Each point in the graph denotes an author with their influence and Topic spread Score. Color of the points are mapped to the logarithm of their aggregate score they have received on their comments and the size of the point is mapped to the logarithm of the number of comments. The graph shows that users with higher influence have higher topic spread and clearly stand out from the rest as red and yellow points.
Hence the proposed measures are able to identify the experts correctly.

\begin{figure}[h]
    \centering
    \includegraphics[width=\linewidth]{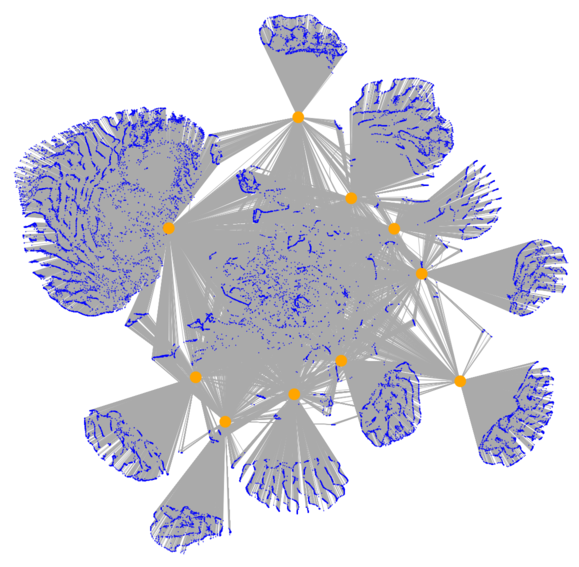}
    \caption{Interaction Network of Top 10 Influencers}
    \label{fig:network}
\end{figure}

Fig.~\ref{fig:network} depicts the network of the top 10 influencers. It is seen that, by and large, each influencer has a community built around them. The communities are not totally isolated, but largely disconnected from each other.  

\section{Related Work}
\label{sec:relatedwork}

\textbf{Behavioral Finance, Social Media and Social Network Analysis}:
Emotion and other factors often influence our decisions. Behavioural Finance is the study of investors’ psychology while making financial decisions. This involves analysis of market level data, like returns and trading volume to study trading behavior of investors. Social media captures the activity of individuals, interactions among them, or more precisely, the complex behavior of a society.  There are studies on Twitter~\cite{oliveira2017impact} that attempt to predict returns, volatility, trading volume and survey sentiment indices -- suggesting that mining social media may provide valuable actionable intelligence to investors. Another study suggest that differences in personalities drive decision making in different ways~\cite{sattar2020behavioral}. Individuals interact with each other in many ways but determining how they interact and uncovering the function of social patterns can be done using social network analysis. A recent review on social network analysis can be found in Ref.~\cite{froehlich2021social}.

\textbf{Topic Modelling:}
LDA has been used predominantly in finding the discussion topics in text data. One such work on publicly available Twitter Data~\cite{zogan2021depressionnet}  uses LDA to do behavior modeling. A large body of work at CLEF eRisk track~\cite{losada2020erisk} aims at detecting mental disorders at an early stage using Reddit data. The use of language markers, topic models, emotion analysis, etc. in social media data has been surveyed in Ref.~\cite{rissola2021survey}. In both of these~\cite{yin2020detecting,prabhakar2020informational}, topics were extracted from Tweets, and their sentiment analysis was done using VADER. LDA was reported to find the most relevant and accurate topics. Another recent study that focuses on finding the discussion topics in the finance domain using Reddit data~\cite{karpenko2021study}. It used RAKE to find the keywords and then distributed these keywords into topics. They have very small data set to study for this domain as compared to ours which is for a complete year.

\section{Conclusion} 
\label{sec:conclusion}

Social discussion platforms harbour a variety of behavior of the authors involved, through their opinions, concerns and points of view as well as the interactions through discussions with other authors. In this work, specifically focusing on the finance domain, we are able to systematically uncover various layers of information. The analysis of content and interaction of a large repository from a social discussion platform (Reddit, in our case) was carried out. The type of information extracted from such large scale social media data can reveal potentially unknown, rare and crucial insights, which are beyond the scope of surveys.
We have proposed a set of measures which are able to map the user-generated content to the components of behavioral finance, and thus offer insights about user concerns, how they are resolved and by whom.
Additionally, using graph analysis, we are able to get insights about how communities are formed and trust is built.
The high number of votes received by comments 
shows the multiplicative effect of the platform, thereby showing that invisible users are also getting benefited.

In all, our study not only uncovers the financial behavior but also attempts to analyze how knowledge on financial behavior is disseminated over a social network.
This study can be extended to further textual analysis of relationships between users, in terms of financial instruments as well as specific asks, and also for identification of individual expertise on various sub domains of finance. In future, our intent is to further analyze how specific financial behaviors like risk taking, entrepreneurial attitude etc. are shaped up by social media platforms.  
Although this work focuses on finance related discussions, our methodology can, in principle, be extended and adapted to other domains.

\bibliographystyle{IEEEtran}
\bibliography{references}
\end{document}